\documentclass[11pt]{article}
\usepackage{amsfonts, amssymb, amsmath, graphicx, epsfig, lscape, subfigure}

\def\inst#1{$^{#1}$}

  {\hspace*{\fill}$\Box$\par\vspace{4mm}}
  {\hspace*{\fill}$\Box$\par\vspace{4mm}}
  {\hspace*{\fill}\par\vspace{4mm}}

\newcommand{\qed}{$\Box$}
\newcommand{\UBu}{UBm_{11}}
\newcommand{\UBz}{UBm_{10}}

\newcommand{\DHn}{D^H_G(n_1)}

\newcommand{\phik}{\phi(k)}

\begin {document}

\title{Rich-Club Ordering and the Dyadic Effect: Two Interrelated Phenomena}

\author{%
Matteo Cinelli\footnote{Corresponding Author} \and Giovanna Ferraro \and Antonio
Iovanella%
}


\maketitle

\begin{center}
{\footnotesize 
\inst{1} Department of Enterprise Engineering\\

University of Rome ``Tor Vergata''\\
Via del Politecnico, 1 - 00133 Rome, Italy.\\
\texttt{matteo.cinelli@uniroma2.it\\
giovanna.ferraro@uniroma2.it\\
antonio.iovanella@uniroma2.it\\}}
\end{center}

\date{}

\maketitle

\begin{abstract}
Rich-club ordering and the dyadic effect are two phenomena observed in complex networks that are based on the presence of certain substructures composed of specific nodes. Rich-club ordering represents the tendency of highly connected and important elements to form tight communities with other central elements. The dyadic effect denotes the tendency of nodes that share a common property to be much more interconnected than expected. In this study, we consider the interrelation between these two phenomena, which until now have always been studied separately. We contribute with a new formulation of the rich-club measures in terms of the dyadic effect. Moreover, we introduce certain measures related to the analysis of the dyadic effect, which are useful in confirming the presence and relevance of rich-clubs in complex networks. In addition, certain computational experiences show the  usefulness of the introduced quantities with regard to different classes of real networks.

\end{abstract}

\section{Introduction}
\label{Intro}

Complex networks are a valid framework of representation and way to analyze a wide variety of real world phenomena. Thanks to a rapid surge of interest in the field, a set of measures able to capture and quantify the peculiarities of various heterogeneous networks have been subsequently introduced, in order to verify empirical observations at different levels of detail. 
The majority of these measures focus on particular tendencies (dynamics) in nodes linkage and the related emerging phenomena such as assortative mixing~\cite{lusseau2004identifying,newman2003mixing, noldus2015assortativity}, homophily~\cite{McPherson2001}, and the rich-club phenomenon~\cite{colizza2006detecting, zhou2004rich}.

The \textit{rich-club phenomenon} has been defined as the tendency of nodes with a high centrality (usually degree) to form highly interconnected communities. It was initially introduced in~\cite{zhou2004rich} in order to analyze the Internet topology at the Autonomous Systems (AS) level and to provide a reasonable explanation as to why such kinds of networks include tightly interconnected hubs. 

Commonly recognized as a power-law network~\cite{BA1999}, the Internet highlights one of the main properties of such degree distributed networks; i.e. the presence of a small amount of nodes having a surprisingly large number of links, the \textit{rich nodes}.
Defining the rich-club coefficient $\phi(k)$, Zhou and Mondragon~\cite{zhou2004rich} revealed that the Internet topology displays the rich-club phenomenon, and that such tendency can not be completely expressed by regular models that generate power-law degree distribution such as the BA-model~\cite{BA1999}, the fitness model~\cite{bianconi2001competition}, and the Inet-3.0 Model~\cite{winick2002inet}.
Furthermore, they noted that well known network measures, like degree assortativity, cannot detect this phenomenon (for instance some degree disassortative networks show the rich-club phenomenon). For this reason, studies that aim to recognize the connection patterns among rich nodes are crucial, in order to provide important insights into a number of  network structures ranging from computer to chemical and social sciences \cite{valverde2007self, zhou2004rich}. 

This topic gained additional interest in the complex networks literature~\cite{jiang2008statistical, mcauley2007rich,xu2010rich, zlatic2009rich} being applied in the analysis of other structural properties, as well as in more recent studies concerning the structural connectivity in the brains of both animals and human beings~\cite{bullmore2009complex, harriger2012rich, sporns2004organization, zamora2011exploring}.
Among all the contributions, the work of Colizza \textit{et al}.~\cite{colizza2006detecting} is particularly noteworthy, as it utilized null models to analyze the implications of such a phenomenon from a more appropriate quantitative point of view.
The introduction of null models, obtained through a process of network rewiring which preserves the degree sequence~\cite{bansal2009exploring, maslov2002specificity}, was very important in order to overcome a bias when considering the rich-club coefficient $\phi(k)$ alone. Indeed, they proved that all networks, even random ones, share a monotonic increasing behavior of the rich-club coefficient with increasing value of degree. This problem was attributed to an intrinsic feature of every network structure in which \textit{hubs}, i.e. high degree nodes, show a higher probability to be more interconnected than low degree ones.
Essentially, the introduction of null models normalizes the rich-club coefficient $\phi(k)$ dividing it by another coefficient $\phi(k)^{ran}$, which is computed in the same way of $\phi(k)$ but results in an average over a proper set of random networks.

When considering the rich-clubs in networks, one can state that they show, in a broad sense, assortative mixing among club members and, at the same time, disassortative mixing since the hubs are connected even to a certain amount of lower degree nodes. 
In such trend, the rich-club coefficient may be seen as a measure able to quantify the level of assortative mixing either to degree or to a discrete characteristic, i.e. to be part of the club or not.
However, we ignore \textit{a priori} the size of the rich-club and, as mentioned before, the study of assortative mixing using standard measures~\cite{newman2003mixing} is not sufficient in this case.
For this reason, in order to make an association between the two phenomena, and thus meet the recent needs of network science~\cite{watts2017should} by gaining new insights into rich-club ordering from an assortative mixing perspective, other approaches are required.

Within this context, a particularly useful and relevant method was initially introduced by Park and Barab\'asi~\cite{park2007distribution}, who investigated assortative mixing using the so-called \textit{dyadic effect}, which can be observed through a binary characterization of the nodes in a network. The dyadic effect is present when the number of nodes sharing a common characteristic (to be part of the club or not in this case) is larger than expected if the characteristic were distributed randomly on the network. This effect, quantified by proper measures, has been widely appreciated and adopted, with a number of extensions throughout many different contexts \cite{CFI2016, di2012protein, FIP2016, hu2015functional, jiang2010towards}.

In this paper we consider the rich-club phenomenon and the rich-club coefficient as introduced in~\cite{zhou2004rich}, studying them with the solid theoretical basis provided by~\cite{colizza2006detecting}. We contribute with a novel formulation of $\phi(k)$ in terms of dyads and dyadic effect. Using this formulation, we reveal certain related measures which provide alternative and additional tools to confirm and test the presence of rich-club ordering in complex networks.

In more detail, we refine the denominator of the original rich-club coefficient, using a theoretical formulation of the upper bounds to the occurrence of dyads types introduced in \cite{cinelli2017structural}. In addition, we quantify the improvement in the denominator of the rich-club coefficient through a measure that we call $\delta(k)$. We then prove how this new formulation maps into the coefficient $\rho(k)$ introduced by~\cite{colizza2006detecting}, which needs to be considered when the significance of the rich-club coefficient is tested.

Moreover, we compute a novel index called $\overline{\phi}(k)$, complementary to $\phi(k)$, which is able to report the extent and way in which nodes of the club tend to connect to the outside, i.e. to non-club nodes. Using $\overline{\phi}(k)$, we are able to understand whether nodes within the club preferentially connect to other high degree nodes, or to those with lower degree. Additionally, with the procedure shown in~\cite{colizza2006detecting}, we compute the normalized index $\overline{\rho}(k)$, which will be useful for confirming and strengthening the presence of rich-club ordering from a different perspective. 
	
As this approach shows how the rich-club phenomenon and the dyadic effect map into one another, we are able to quantify their magnitude using degree sequence, i.e. combinatorial, based arguments that take into account the different kinds of connections among network nodes.

The paper is organized as follows: Section~\ref{probset} provides the theoretical background; Section~\ref{ref_nor} shows the reformulation and normalization of the rich-club coefficient; Section~\ref{CR} contains the computational analysis and Section~\ref{conclusion} presents the conclusions.

\section{Problem settings}\label{probset}
\subsection{Theoretical Background}
\label{ThBg}

A network can be represented as a graph $G$ with $N$ nodes and $M$ edges and herein we consider each graph as undirected, unweighted, connected and simple, i.e. loops and multiple edges are not allowed.

The degree $d_i$ of a node $i$ is defined as the number of edges incident to $i$. The list of nodes degrees in non-increasing order is called \textit{degree sequence} $D_G$ and for every connected graph holds the Degree-Sum Formula or \textit{Handshaking Lemma},  $\sum_{i=1}^{N}{d_i} = 2M$. A \textit{graphic sequence} is defined as a list of nonnegative numbers, which is the degree sequence of certain simple graphs, and a graph $G$ with degree sequence $D_G$ is called a \textit{realization} of $D_G$.
A generic list $L$ of nonnegative numbers is not necessarily a graphic sequence. Indeed, a necessary and sufficient condition is that $\sum_{i=1}^{N}{d_i}$ is even and  $\sum_{i=1}^{N}{d_i} \leq k(k-1)+ \sum_{i=k+1}^n \min\{k, d_i\}$, $1 \leq k \leq N$ ~\cite{EG1960}. The problem of determining whether $L$  is a graphic sequence is referred to as the \textit{Graph Realization Problem} \cite{hakimi1962realizability, havel1955remark}.

Given an integer $n \leq N$, we consider the subsequence of the first $n$ elements of $D_G$ whereby $D_G^H(n) \subseteq D_G$ is the {\it head} of $D_G$, and the subsequence of the last $n$ elements of $D_G$ whereby $D_G^T(n) \subseteq D_G$ is the {\it tail} of $D_G$.

A \textit{clique} $K_n$ is a complete subgraph of $G$ of dimension $n$, i.e. a subgraph of $n$ mutual interconnected nodes. A \textit{star} $S_n$ is a subgraph of $G$ of dimension $n$ showing one node with degree $n-1$ and the others $n -1$ having degree $1$.

\subsection{The rich-club coefficient}
\label{rich_coeff}

For our purpose, in this paper we consider the rich-club coefficient expressed in terms of nodes degree as in~\cite{colizza2006detecting} that can be written as:

\begin{equation}
\phi(k) = {\frac {2 E_{>k}}{N_{>k}(N_{>k}-1)}}
\label{phi}
\end{equation}

\noindent where $E_{>k}$ is the number of edges among the $N_{>k}$ nodes having degree $d_{i}$ higher than a given value $k$ and $\frac {N_{>k}(N_{>k}-1)}{2}$ is the maximum possible number of edges among the $N_{>k}$ nodes. Therefore $\phik$ measures the fraction of edges connecting the $N_{>k}$ nodes out of the maximum number of edges they might possibly share.

\subsection{Nodes characteristics and the dyadic effect}
\label{nodechar}

\begin{figure}[htbp]
\begin{center}
\includegraphics[trim=3.5cm 24.3cm 3cm 3cm, clip = true, totalheight=0.08\textheight]{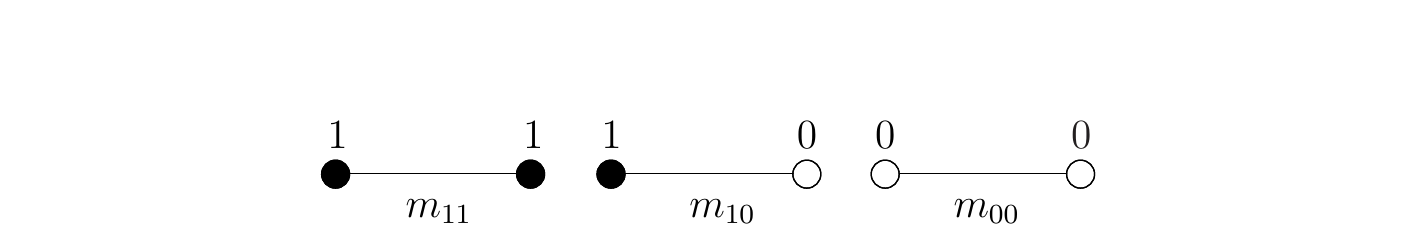}
\caption{Types of dyads.}
\label{bpd}
\end{center}
\end{figure}

Herein, we refer to a given characteristic $c_i$, which can assume the values $0$ or $1$, for each $i \in N$. Consequently, $N$ can be divided into two subsets: the set of $n_1$ nodes with characteristic $c_i = 1$, the set of $n_0$ nodes with characteristic $c_i = 0$; thus, $N = n_1 + n_0$. We distinguish three kinds of \textit{dyads}, i.e. edges and their two end nodes, in the network: $(1 -� 1)$, $(1 -� 0)$, and $(0 -� 0)$ as depicted in the Figure~\ref{bpd}.

We label the number of each dyad in the graph as $m_{11}$, $m_{10}$, $m_{00}$, respectively. Hence, $M = m_{11} + m_{10} + m_{00}$. We consider $m_{11}$ and $m_{10}$ as independent parameters that represent the dyads containing nodes with characteristic 1.

\begin{figure}[htbp]
\begin{center}
 \begin{minipage}[b]{6cm}
   \centering
   \includegraphics[scale=.8]{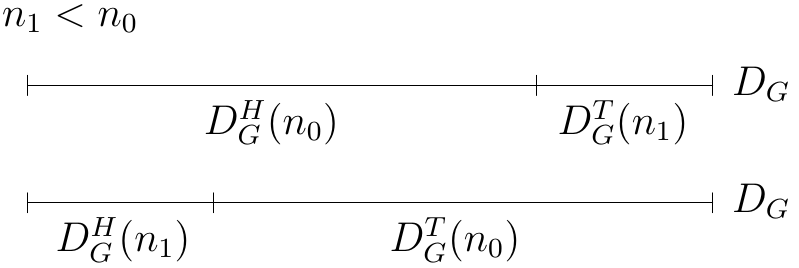}
   \caption{Two different partitions when $n_1 < n_0$.}\label{deg1}
 \end{minipage}
 \ \hspace{5mm} \hspace{5mm} 
 \begin{minipage}[b]{6cm}
  \centering
   \includegraphics[scale=.8]{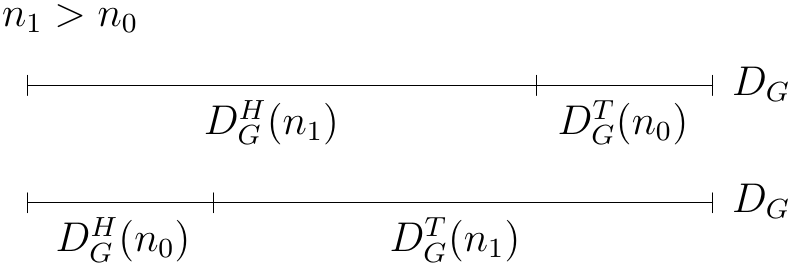}
   \caption{Two different partitions when $n_1 > n_0$.}\label{deg2}
 \end{minipage}
 \end{center}
\end{figure}

Let $D_G$ be the degree sequence of $G$, we can use $n_1$ and $n_0$ to define its head $D_G^H(n_1)$ or $D_G^H(n_0)$ and its tail $D_G^T(n_1)$ or $D_G^T(n_0)$ such that $D_G = D_G^H(n_1) \cup D_G^T(n_0)$ or  $D_G = D_G^H(n_0) \cup D_G^T(n_1)$. These partitions of the degree sequence are given arbitrarily, assigning the characteristic $c_i = 1$  to the $n_1$ nodes with the highest degree or to the $n_1$ nodes with the lowest degree or viceversa. Such partitions are reported in Figure~\ref{deg1} and~\ref{deg2}, distinguishing the case in which $n_1 < n_0$ or $n_1 > n_0$.

When nodes in a network fit within two distinct groups according to their characteristics, two different parameters are required to determine the existence of the relations between the network topology and the nodes features~\cite{park2007distribution}. Whether the number of edges between nodes sharing a common characteristic is larger than expected if the characteristics are distributed randomly on the graph we observe a  phenomenon called the {\it dyadic effect}~\cite{park2007distribution}.
Considering the $N$ nodes and the case in which any node possesses an equal chance of having the characteristic 1, given $n_1$, the expected values of $m_{11}$ and $m_{10}$ are~\cite{park2007distribution}:

\begin{equation}
\overline{m}_{11} = {n_1 \choose 2} \delta = \frac{n_1 (n_1 - 1)}{2} \delta
\label{m11}
\end{equation}

\begin{equation}
\overline{m}_{10} = {n_1 \choose 1} {n_0 \choose 1} \delta = n_1 (N - n_1) \delta
\label{m10}
\end{equation}

\noindent
where $\delta$ is the density and is equal to $\delta = 2M/N(N - 1)$. The relevant deviations of $m_{11}$ and $m_{10}$ from the expected values 
$\overline{m}_{11}$ and $\overline{m}_{10}$ denote that the characteristic $1$ is not randomly distributed~\cite{deAlmeida2013, park2007distribution}. Such deviations can be calculated through the ratios called \textit{dyadicity} $D$ and \textit{heterophilicity} \textit{H} defined as: 
\begin{equation}
D = \frac{m_{11}}{\overline{m}_{11}}
\label{eq_D11}
\end{equation}

\begin{equation}
H = \frac{m_{10}}{\overline{m}_{10}}
\label{eq_H}
\end{equation}


In~\cite{park2007distribution}, it is established that $m_{11}$ and $m_{10}$ cannot assume arbitrary values, as there are indirect constraints due to the network structure. Indeed, $m_{11}$ cannot exceed 
\begin{equation}\label{ub11old}
\UBu = min( M, \binom { n_1 } {2} )
\end{equation}
\noindent
and $m_{10}$ cannot be larger than 
\begin{equation}\label{ub10old}
\UBz = min( M, n_1 n_0)
\end{equation}

\noindent where \textit{UB} stands for upper bound. These upper bounds are improved in \cite{cinelli2017structural}, which explicitly outlines the structural constraints regarding the $m_{11}$ and $m_{10}$ links.  The new upper bounds, which are herein used throughout the paper, are written as:

\begin{equation}\label{for_m11}
UBm_{_{11}} = min\Bigg( M, \binom { n_1 } {2}, \bigg\lceil \sum_{i \in \DHn}{\frac {\min(d_{i}, n_1 - 1)}{2}} \bigg\rceil \Bigg)     
\end{equation}

\begin{equation}\label{for_m10}
UBm_{10} = min\Bigg( M, n_1 n_0, 
min\bigg( \sum_{i \in D_G^H(n_1)}{\min(d_{i}, n_0)} , \sum_{i \in D_G^H(n_0)}{\min(d_{i}, n_1)}  \bigg)\Bigg)     
\end{equation}

The first upper bound is based on the fact that large cliques may be rare substructures in networks~\cite{bianconi2006emergence, bollobas1998random}; therefore, using $D_G$ we check whether $G$  can actually contain a complete subgraph of size $n_1$. If not, we take into account the densest hypothetical substructure that could be realized using the degree sequence of $G$. In the second upper bound we check if $G$ can contain $n_1$ stars of size $n_0$ or $n_0$ stars of size $n_1$ looking at the degree sequence $D_G$. If not, we consider a set of stars made of the first $n_1$ elements of $D_G$ if $n_1 < n_0$ or made of the first $n_0$ elements of $D_G$ if $n_0 < n_1$.
The exhaustive explanations and proofs are described in \cite{cinelli2017structural}.

\section{Reformulation and normalization of the rich-club coefficient}\label{ref_nor}

In this section we provide a reformulation of the rich-club coefficient involving most of the measures used in the study of the dyadic effect.

\subsection{Reformulation of the rich-club coefficient}
\label{ref_of_rcc}

Let us write:
 \[
 c_{i} = 
  \begin{cases} 
   1 & \text{if } d_{i} > k \\
   0       & \text{if } d_{i} \leq k
  \end{cases}
\]
we immediately obtain $N_{>k} = n_1$ and $N_{\leq{k}} = n_0$. Since $E_{>k}$ is defined as the number of edges between nodes $N_{>k} = n_1$, we can consequently write $E_{>k} = m_{11}$ while the maximum possible number of links between nodes $N_k$ can be written as $\frac {N_{>k}(N_{>k}-1)}{2} = \binom{n_1}{2}$.

Let us also define $\overline{E}_{>k} = M - E_{>k} - E_{\leq{k}}$ as the number of edges among the $N_{>k}$ and $N_{\leq{k}}$ nodes having degree $d_i$ greater and less equal than a given value $k$. Note that $N_{>k} N_{\leq{k}}$ is the maximum possible number of edges among the $N_{>k}$ and $N_{\leq{k}}$ nodes and, given that we have a nodal bipartition, thus $N = N_{>k} + N_{\leq{k}} = n_1 + n_0$.
Therefore, we can write $N_{>k} N_{\leq{k}} = n_1 n_0$ and, using the equality $\overline{E}_{>k} = M - E_{>k} - E_{\leq{k}} = M - m_{11} - m_{00} = m_{10}$ thus $\overline{E}_{>k} = m_{10}$.
A summary of the discussed measures is reported in Table~\ref{table_n1}.

\begin{table}[b]
 \label{table_n1}
\centering
    \begin{tabular}{| r || c | c | c | c | c | c |}
    \hline
    Rich-club notation & $N_{>k}$ & $N_{\leq{k}}$ & $E_{>k}$ & $\overline{E}_{>k}$ & $E_{\leq{k}}$ \\ \hline
    Dyadic effect notation & $n_1$ & $n_0$ & $m_{11}$ & $m_{10}$ & $m_{00}$ \\ \hline
    \end{tabular}
\caption{Correspondences between rich-club measures and their counterparts in terms of dyadic effect.}
\end{table}

 Under these circumstances we can reformulate $\phik$ as:
\begin{equation}
\phi(k) = {\frac {2 E_{>k}}{N_{>k}(N_{>k}-1)}} = {\frac {m_{11}}{\binom {n_1}{2}}}
\label{phi_new}
\end{equation}
As previously outlined, by knowing the degree sequence $D_G$, we understand that the maximum number of links between the $n_1$ nodes is bounded by $UBm_{_{11}}$ and, as shown in \cite{cinelli2017structural}, it can be much lower than $\binom{n_1}{2}$.
Through this upper bound, we observe if the considered graph, as well as all the other ones having the same $D_{G}$, can contain a complete subgraph of order $n_1$. If such case does not occur, $D_{G}$ can not generate a graph containing a clique with $n_1$ nodes and we can compute the rich-club coefficient obtaining a more realistic value, tailored on the degree sequence of the graph under consideration.

This approach may lead to higher values of $\phik$, and thus a better understanding and quantification of the phenomenon, since we are taking into account a proper measure of dyadic effect. Indeed, in this case, the denominator of the coefficient embeds an informative content which derives directly from a structural property of the network, i.e. the degree sequence $D_{G}$. Finally we can reformulate an improved version of $\phik$ as:
\begin{equation}
\phi(k)^{new} = {\frac {m_{11}}{UBm_{_{11}}}}
\label{phi_new_1}
\end{equation}

\subsection{The complementary rich-club coefficient}
\label{compl_rcc}

While computing $m_{11}$ in order to have $\phik$ or $\phi(k)^{new}$ we can easily compute $m_{10}$, i.e. the number of links between the nodes of the rich-club ($n_1$) and the nodes outside the club ($n_0$). 
Using this approach, we are able to quantitatively determine to which extent the nodes of the rich-club are connected to others in the network by making the following comparison: which portion of the total degree of the club members realizes internal connections, i.e. $m_{11}$, versus which portion realizes external connections, i.e. $m_{10}$. 
Under these circumstances, it is useful to exploit a classification of network links that has been provided (for instance, in~\cite{van2012high}) in order to identify different classes of links within the human brain in case of rich-club ordering. In the article, the different kinds of links are called \textit{rich-club connections} if they link rich-club nodes, \textit{feeder connections} if they link rich-club nodes to nonrich-club nodes and \textit{local connections} if they link nonrich-club nodes to each other. In our notation the rich-club connections correspond to $m_{11}$, the feeder connections correspond to $m_{10}$ and the local connections correspond to $m_{00}$.


Following the same rationale as before, we can quantify the interaction among club and non-club members through the index $\overline{\phi}(k) = \frac {\overline{E}_{>k}}{N_{>k} N_{\leq{k}}}$ and \textit{ipso facto} reformulate and improve it with certain considerations similar to those of Section \ref{ref_of_rcc}. 

Summarizing we can initially write $\overline{\phi}(k) = \frac {m_{10}}{n_1 n_0}$ and, using the formulation of $UBm_{_{10}}$ as mentioned in Section~\ref{ThBg}, express $\overline{\phi}(k)$ as:
\begin{equation}
\overline{\phi}(k) = {\frac {m_{10}}{UBm_{_{10}}}}
\label{phi__segn_new}
\end{equation}

The index $\overline{\phi}(k)$ represents the number of feeder connections divided by the maximum possible number of such connections. 
\subsection{On the normalization of the rich-club coefficient}
\label{norm_of_rcc}

The use of null models and of the normalization process of topological and structural measures in complex networks represents a practice widely used to comprehend the real magnitude of certain phenomena. For this reason, the normalization of the rich-club coefficient, suggested in~\cite{colizza2006detecting} and adopted in many further studies~\cite{bullmore2009complex, harriger2012rich, jiang2008statistical, mcauley2007rich, sporns2004organization, zamora2011exploring, zlatic2009rich}, is a necessary procedure that has to be adopted in order to take into account the significance of this index.
The procedure involves an ensemble of random networks, sometimes called \textit{maximally random networks}~\cite{colizza2006detecting, mcauley2007rich}, which have the same degree distribution and sequence as the one under investigation that, if generated in a sufficiently large number, provide a null distribution of the rich-club coefficient. 

Using this method, the normalized rich-club coefficient is defined~\cite{colizza2006detecting} as: 
\begin{equation}
{\rho}(k) = {\frac {\phi(k)}{\phi(k)_{ran}}}
\label{rho}
\end{equation}

\noindent where $\phi(k)_{ran}$ is the average rich-club coefficient across the random network ensemble. We observe rich-club ordering when $\rho(k) > 1$. As previously mentioned, the process generates random networks with prescribed degree sequence $D_{G}$, thus the number of nodes with degree higher than $k$, $N_{>k}$, will be the same for each $k$ and for each network with the same $D_{G}$ of the original one, implying that the denominator of $\phi(k)$, i.e. $\frac{N_{>k} (N_{>k} - 1)}{2}$, will be the same for either $\phi(k)$ or $\phi(k)_{ran}$. We can consequently write:  
\begin{equation}
{\rho}(k) = {\frac {\frac{2 E_{>k}}{N_{>k} (N_{>k} - 1)}}{{\frac{2 E_{>k}|_{ran}}{N_{>k} (N_{>k} - 1)}}}} = \frac {E_{>k}}{E_{>k}|_{ran}}
\label{rho_simple}
\end{equation}

or using the dyadic effect notation:

\begin{equation}
{\rho}(k) = {\frac {m_{11}}{m_{11}|_{ran}}}
\label{rho_dyad}
\end{equation}

\noindent where $m_{11}|_{ran} = E_{>k}|_{ran}$ is the average value of $m_{11}$ across the random networks ensemble.
Similar considerations hold even when we take into account $\phi(k)^{new}$ and they are motivated by the nature of its denominator, $UBm_{11}$, that depends uniquely on the degree sequence $D_{G}$ and includes the quantity $\binom{n_1}{2} = \frac{N_{>k} (N_{>k} - 1)} {2}$, as shown in \cite{cinelli2017structural}. Therefore, for any value of $k$, $UBm_{11}$ will be the same for each network, having the same degree sequence $D_{G}$, leading to:

\begin{equation}
{\rho}(k)^{new} = {\frac {\frac {m_{11}}{UBm_{11}}}{\frac {m_{11}|_{ran}}{UBm_{11}}}} = {\frac {m_{11}}{m_{11}|_{ran}}}
\label{rho_new}
\end{equation}

\noindent and subsequently to the equality $\rho(k)^{new}$ = $\rho(k)$.

This implies that our approach improves the rich-club coefficient $\phi(k)$ itself, and that the evaluation of the presence of such ordering maps into the well known approach of~\cite{colizza2006detecting}. So when we observe the presence of the rich-club ordering, i.e $\rho(k)>1$ and $\phik \neq 0$, we can evaluate its magnitude by observing the value of $\phi(k)^{new}$ or more precisely compare it with the value of $\phi(k)$ considering the coefficient $\delta(k)$ defined as:

\begin{equation}
\delta(k) = {\frac {\phi(k)^{new} - \phi(k)}{\phi(k)^{new}}}
\label{eval}
\end{equation}

Since $\phi(k)^{new} \geq \phi(k)$ for each $k$, $\delta(k) \in [0,1)$. 

The coefficient $\delta(k)$ may be helpful in further testing the significance of rich-club ordering in networks. It is important to underline that for high values of $k$, the number of nodes within the rich-club decreases, as does the denominator of $\phi(k)$. In this case, the value $UBm_{11}$ tends to equal the number of links within the clique, leading to $\delta(k)$ being valued at zero. The index $\delta(k)$ underlines the fact that very large cliques are uncommon in real networks \cite{bianconi2006emergence} and that the presence of dense substructures should be conducted using measures, such as $UBm_{11}$, that are more related to the network under investigation. 

When $\delta(k) > 0$ then $\phi(k) \leq \phi(k)^{new}$ because $\binom{n_1}{2} \geq UBm_{11}$. Indeed, $UBm_{11}$ exploits combinatorial arguments and captures the densest realizable substructure when a clique of $n_1$ nodes is not realizable within the given $D_G$. Thus, $\phi(k)^{new}$ is more likely to identify the best (densest) rich-club not as a clique, but as a sparser subgraph of the same size. 

Another relevant information derives from the analysis of the relationships between club and non-club members, i.e. from the so-called feeder connections, using the aforementioned coefficient $\overline{\phi}(k)$. It compares the number of feeder connections in the network with the number of feeder connections across the random network ensemble, quantifying the portion of edges that have start point in club nodes and end point in non-club nodes. Even in this case, we are able to adopt the normalization process to obtain a second coefficient: $\overline{\rho}(k)$. This can be helpful in capturing and evaluating the interplay among different classes of nodes, quantifying for each club member the portion of its degree invested in the outside and computing the same quantity across all the club members. Therefore, we can write 
\begin{equation}
\overline{\rho}(k) = {\frac {\overline{\phi}(k)}{\overline{\phi}(k)_{ran}}}
\label{}
\end{equation}

where $\overline{\phi}(k)_{ran}$ is the average across the random network ensemble. 

For networks showing rich-club ordering we expect that, compatibly with the size of the club, most of the degree of club members will be invested in links with other club members, meaning ${\overline{\phi}(k)} \leq {\overline{\phi}(k)_{ran}}$. In other words, when rich-club ordering is present, we expect members showing $d_{i} > k$ ( i.e. $n_1$) to be more connected among themselves that to other with $d_{i} < k$ (i.e. $n_0$) especially in comparison to randomized networks. 

Thus, with the presence of rich-club ordering, it is expected that the curve of $\overline{\rho}(k)$ approaches the value $\overline{\rho}(k)=1$ from below while $k$ increases, meaning that we have fewer feeder connections than in random networks having the same degree sequence as that of the original. Indeed, $m_{10} \leq m_{10}|_{ran}$ means that high degree nodes tend to connect more densely to one another than to low degree nodes. 

It is also important to point out that the trend of the curve $\overline{\rho}(k)$ reaching the value $\overline{\rho}(k) = 1$ (either from below or from above) is not surprising, especially for high values of $k$, since for such values the clique size could be very small. For instance, let us consider a certain network showing rich-club ordering. If we imagine that the club is made up of the three highest degree nodes of the network, a triangle would be the biggest clique, thus allowing the rich-club to have three links at most. In this case, the three nodes will necessarily connect to other ones showing a lower degree, thereby increasing the chance of having ${\overline{\phi}(k)} = {\overline{\phi}(k)_{ran}}$. 
For these reasons, the values assumed by $\overline{\rho}(k)$ have to be considered collectively via a curve that is able to represent the tendency of high degree nodes to connect mainly to other high degree nodes; i.e. a curve able to unveil the tendency of nodes to form connected cores of important members. If the curve $\overline{\rho}(k)$ approaches the value $\overline{\rho}(k) = 1$ from below, this seems to further confirm the presence of rich-club ordering.

\section{Computational Results} 
\label{CR}

Herein, we show computational results for three different networks: the AS-level Internet topology as in~\cite{zhou2004rich}, the top 500 US airports network as in~\cite{colizza2007reaction} and the Scientific Collaboration Network as in~\cite{newman2001structure}. In order to compute $\rho(k)$ for the three case studies we consider a set of 1000
random graphs generated by randomizing the connections of the adjacency matrix while preserving the original degree sequence such as, for instance, in~\cite{colizza2006detecting, collin2014structural, van2011rich}. The data processing, the network analysis and all simulations were performed using the software {\it R}~\cite{R} with the {\it igraph} library~\cite{CN2006}.

\begin{figure}[htbp]
\begin{center}
\includegraphics[scale=.5]{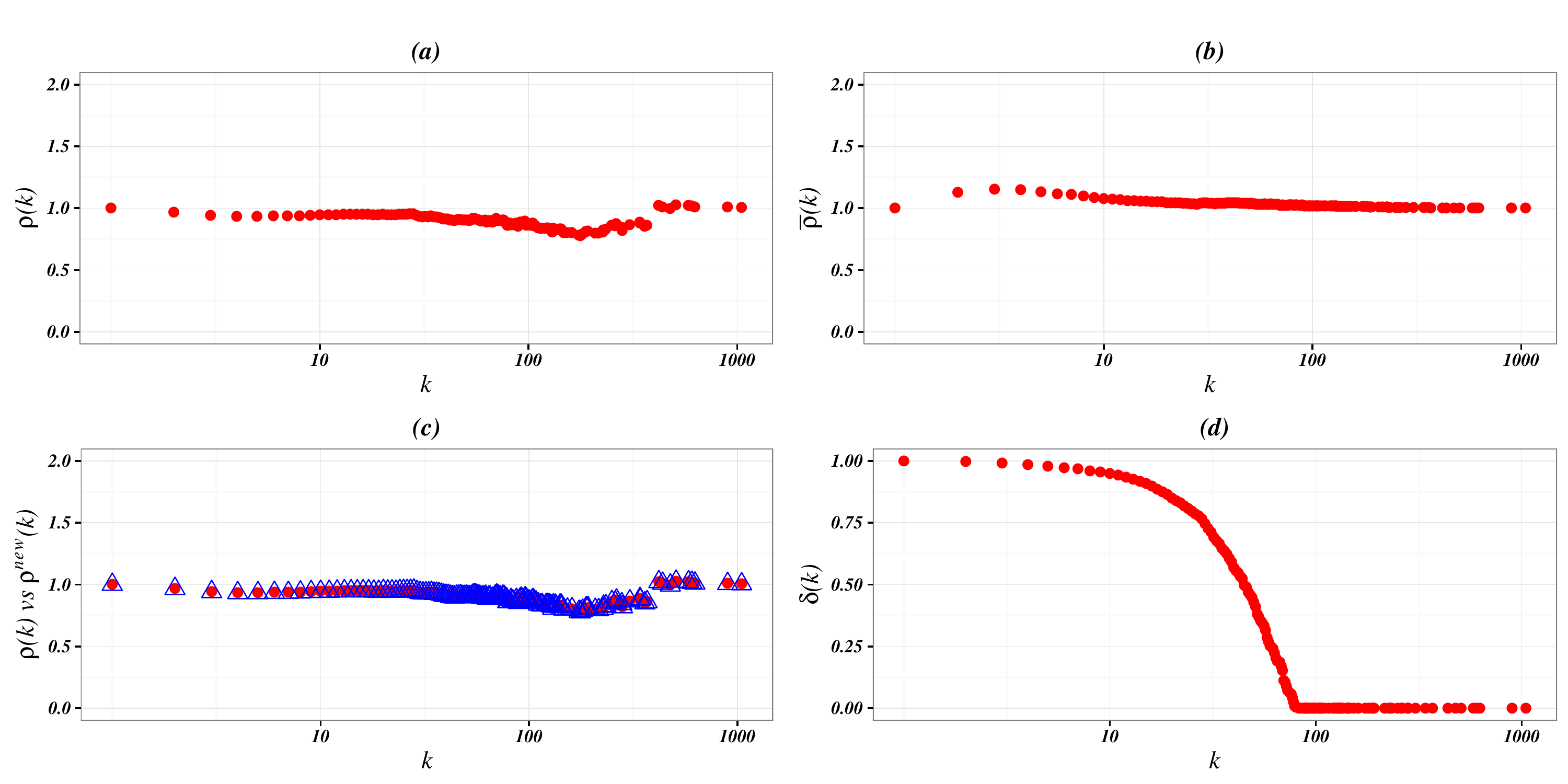}
\caption{Curves related to $\rho(k)$, $\overline{\rho}(k)$, $\rho(k)$ \textit{vs} $\rho(k)^{new}$ and $\delta(k)$ for the  AS-level Internet network.}
\label{Int}
\end{center}
\end{figure}

In the case of the Internet topology and referring to $\rho(k)$ in Figure~\ref{Int} (a), we obtained results corresponding to those of~\cite{colizza2006detecting} that are in some way strengthened by the fact that $\overline{\rho}(k)$ is approaching the value $\overline{\rho}(k) = 1$ from above, as shown in Figure~\ref{Int} (b). This means that the feeder connections are distributed as in random networks with the same degree sequence of the original one. This shows the low tendency of hubs to connect mainly to other hubs, thus confirming the absence of a strong rich-club ordering in such a structure. It is also worth mentioning at this point that the study of the rich-club ordering on the Internet topology is still a debated issue \cite{jiang2008statistical, xu2010rich}.

\begin{figure}[htbp]
\begin{center}
\includegraphics[scale=.5]{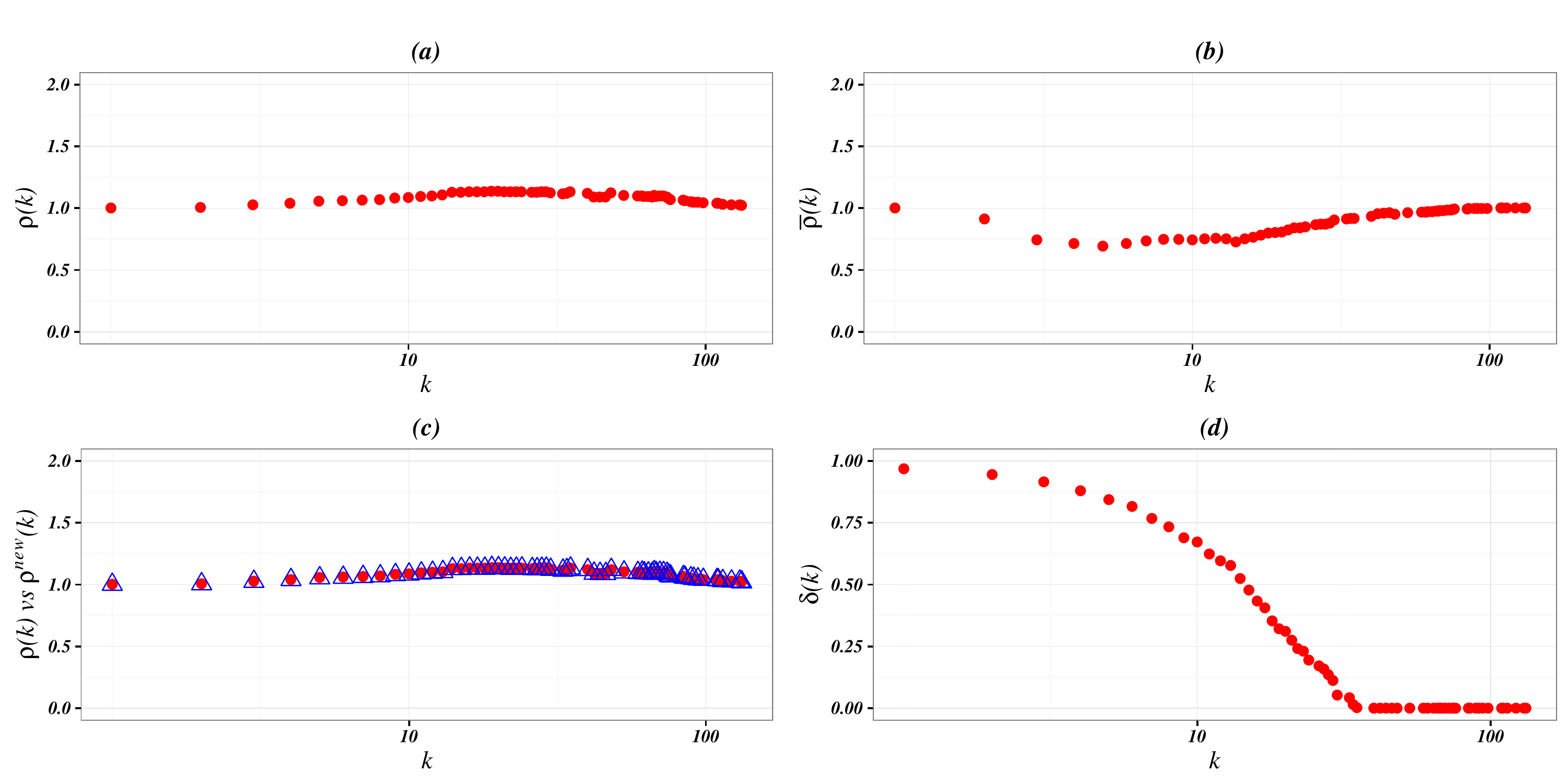}
\caption{Curves related to $\rho(k)$, $\overline{\rho}(k)$, $\rho(k)$ \textit{vs} $\rho(k)^{new}$ and $\delta(k)$ for the top 500 US airports network.}
\label{Air}
\end{center}
\end{figure}

In the case of the top 500 US airports network, we observe a stronger rich-club ordering as shown in  Figure~\ref{Air} (a), which is still confirmed by the distribution of the feeder connections approaching the value $\overline{\rho}(k) = 1$ from below as depicted in Figure~\ref{Air} (b). Indeed, in this network, $\overline{\rho}(k)$ is always less than 1, thus the nodes within the club tend to be more densely connected among themselves than observed in random networks. 

\begin{figure}[htbp]
\begin{center}
\includegraphics[scale=.5]{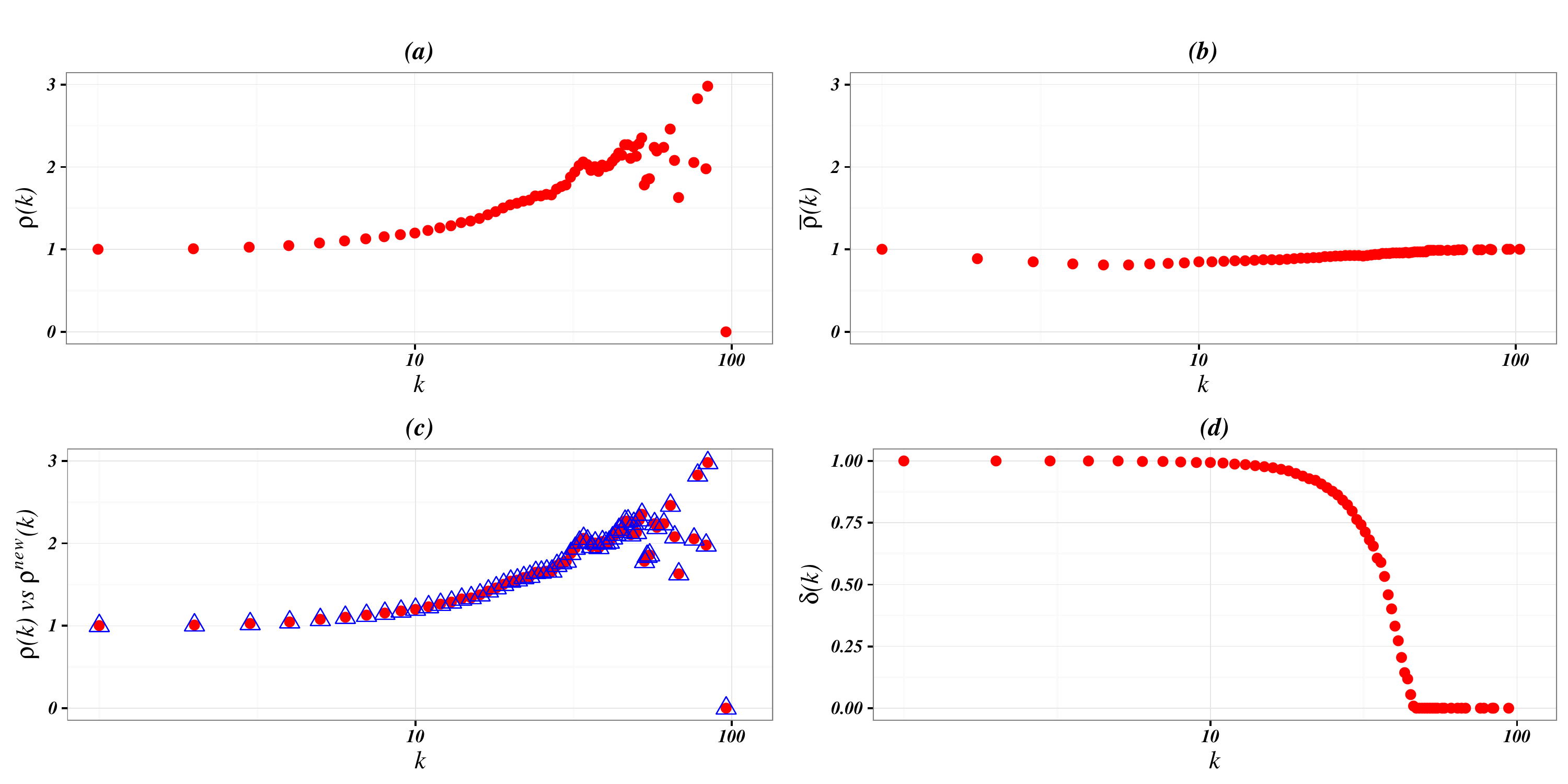}
\caption{Curves related to $\rho(k)$, $\overline{\rho}(k)$, $\rho(k)$ \textit{vs} $\rho(k)^{new}$ and $\delta(k)$ for the Scientific Collaboration network.}
\label{Cit}
\end{center}
\end{figure}

In the third case of Citation network, shown in Figure~\ref{Cit}, we have the strongest rich-club ordering, once again confirmed by the trend of $\overline{\rho}(k)$ approaching the value $\overline{\rho}(k) = 1$ from below. These three examples seem to confirm the discriminating power of $\overline{\rho}(k)$.

Figures~\ref{Int},~\ref{Air} and~\ref{Cit} (c) contain the curves of $\rho(k)$ and $\rho(k)^{new}$, computationally confirming the previous analytical result regarding their equality. Figures ~\ref{Int} and ~\ref{Air} (d) show the values of the coefficient $\delta(k)$ in order to evaluate the gain introduced by the definition of $\phi(k)^{new}$. The trend of the curves mainly depends on the shape of the degree sequence $D_G$ thus, on the degree distribution of the network under examination, since the new upper bound (the denominator of $\phi(k)^{new}$) is computed using that element. The decreasing behavior of the curve is not surprising; indeed while the degree $k$ grows, the number of elements with degree higher than $k$ ($N_{k} = n_1$) decreases, implying a consequent reduction in the number of links within the clique of size $N_k$. When the clique is composed of very few elements, it is likely that the new upper bound $UBm_{11}$ maps into the quantity at the denominator of $\phi(k)$ implying $\delta(k) = 0$.

\section{Conclusions}
\label{conclusion} 

In this paper, we investigated the organization of rich nodes in networks from a slightly different perspective: the dyadic effect. We proved how rich-club ordering and the dyadic effect have much in common, showing how one can map into another and viceversa. Exploiting certain properties concerning the study of dyads, we provided additional instruments in order to investigate and confirm the presence of rich-club ordering. As previously mentioned, the issue of knowing whether a certain network possesses a rich-club is important for many reasons, including factors relating to either its functional or topological role. Indeed, the presence of this peculiar substructure is relevant in problems such as resilience and information flow in networks, mainly because of its influence on crucial properties like assortativity and transitivity.

From the computational point of view, the analysis of rich-club in terms of the dyadic effect needs the same resources of its original formulation since it provides simultaneously (i.e from a single network observation) the values of $m_{11}$ and $m_{10}$ that are necessary in order to obtain $\phi(k)$ and $\overline{\phi}(k)$. The last observation holds even for what concerns the computation of the set of rewired networks that is the same for both $\rho(k)$ and $\overline{\rho}(k)$. 
The consistency of our approach with the literature, in term of normalization processes, represents a further advantage because it includes a more detailed information content related to the presence of both rich-club connections and feeder connections.   

As further investigation, it may be interesting to study the organization of non-rich nodes and their connection patterns. For instance, a coefficient $\widetilde{\phi}(k) = \frac{m_{00}}{UBm_{00}}$ (where $UBm_{00}$ corresponds to $UBm_{11}$ when considering nodes with $c_{i}=0$) may be introduced to test the connectance among nodes linked by local connections. However, the analysis of such an aspect could represent a difficult issue to deal with because of the interpretation given to the presence of highly interconnected nodes with relatively low degree.

Another important aspect may concern the relationship between rich-club ordering and other centrality measures beyond degree. It would be interesting to study dyadic interactions among nodes possessing values of centrality over a certain richness threshold. Lastly, the extension of the dyadic effect to weighted and directed networks could provide room for new developments linked to results in the existing literature that has focused on the investigation of particular networks, such as the World Trade Web and the brain networks in mammals.








\end{document}